\documentclass[journal,10pt]{IEEEtran}

\usepackage{amsfonts}
\usepackage{amsmath}
\usepackage{graphicx}
\usepackage{color}
\usepackage{multicol}
\usepackage{amssymb}
\usepackage{subfigure}
\usepackage{bm}
\usepackage{latexsym}
\usepackage{stfloats}
\usepackage{float}
\usepackage{exscale}
\usepackage{relsize}
\usepackage{cite}
\usepackage{amsthm,amsmath}
\usepackage[ruled,linesnumbered]{algorithm2e}
\usepackage{mathtools}

\begin{document}
\vspace{-0.5cm}
\title{RIS-Aided D2D Communications Relying on Statistical CSI with Imperfect Hardware}

\author{\vspace{-0.1cm}
        Zhangjie~Peng,
        Tianshu~Li,
        Cunhua~Pan,~\IEEEmembership{Member,~IEEE,}
        Hong~Ren,~\IEEEmembership{Member,~IEEE,}
        \\
        and~Jiangzhou Wang,~\IEEEmembership{Fellow,~IEEE}
        % <-this % stops a space
\vspace{-0.2cm}

\thanks{(\emph{Corresponding author: Cunhua Pan}.)}
\thanks{Z. Peng is with the College of Information, Mechanical, and Electrical Engineering, Shanghai Normal University, Shanghai 200234, China, also with the National Mobile Communications Research Laboratory, Southeast University, Nanjing 210096, China, and also with the Shanghai Engineering Research Center of Intelligent Education and Bigdata, Shanghai Normal University, Shanghai 200234, China $( \text{e-mail: pengzhangjie@shnu.edu.cn} )$.
}% <-this % stops a space
\thanks{T. Li is with the College of Information, Mechanical and Electrical Engineering,
Shanghai Normal University, Shanghai 200234, China $( \text{e-mail: 1000479056@smail.shnu.edu.cn} )$.
}% <-this % stops a space
%\thanks{L. Kong is with the H3C Technologies Co., Limited, Hangzhou, Zhejiang 310052, China $( \text{e-mail: kong.lei@h3c.com} )$.}% <-this % stops a space
\thanks{C. Pan is with the School of Electronic Engineering and Computer Science at Queen
Mary University of London, London E1 4NS, U.K. $( \text{e-mail: {c.pan}@qmul.ac.uk} )$.}% <-this % stops a space
\thanks{H. Ren is with the National Mobile Communications Research
Laboratory, Southeast University, Nanjing 210096, China $( \text{e-mail: {hren}@seu.edu.cn} )$.}% <-this % stops a space
%\thanks{W. Xu is with National Mobile Communications Research Laboratory,
%Southeast University, Nanjing 210096, China $( \text{e-mail: wxu@seu.edu.cn} )$.}% <-this % stops a space
\thanks{J. Wang is with the School of Engineering and Digital Arts, University of Kent, CT2 7NT Canterbury, U.K. (e-mail: j.z.wang@kent.ac.uk).}
\vspace{-1cm}
}

\maketitle

\newtheorem{lemma}{Lemma}
\newtheorem{theorem}{Theorem}
\newtheorem{remark}{Remark}
\newtheorem{corollary}{Corollary}

\vspace{-1cm}
\begin{abstract}
In this letter, we investigate a reconfigurable intelligent surfaces (RIS)-aided device to device (D2D) communication system over Rician fading channels with imperfect hardware including both hardware impairment at the transceivers and  phase noise at the RISs.
This paper has optimized the phase shift
by a genetic algorithm (GA) method to maximize the achievable
rate for the continuous phase shifts (CPSs) and discrete phase
shifts (DPSs).
We also consider the two special cases of no RIS hardware impairments (N-RIS-HWIs) and no transceiver hardware impairments (N-T-HWIs). We present closed-form  expressions for the achievable rate of different cases and study the impact of hardware impairments on the communication quality.
Finally, simulation results validate the analytic work.
%In this letter, we investigate a reconfigurable intelligent surface (RIS) aided multi-pair communication system, in which multi-pair users exchange information via an RIS. We derive an approximate \textcolor{blue}{expression for} the achievable rate by assuming that statistical channel state information (CSI) is available. A genetic algorithm (GA) to solve the rate maximization problem is proposed as well.
%In particular, we consider implementations of RISs with continuous phase shifts (CPSs) and discrete phase shifts (DPSs).
%Simulation results \textcolor{blue}{verify} the obtained results and show that the proposed GA method has almost the same performance as the globally optimal solution.
%In addition, numerical results show that three quantization bits can achieve a large portion of the achievable rate for the CPSs setup.
\end{abstract}


\IEEEpeerreviewmaketitle

\begin{figure*}[hb]
\vspace{-0.5cm}
\hrulefill
\vspace{-0.2cm}
\setcounter{equation}{9}
\begin{align}\label{SINR_i}
{\rm{\gamma}}_{i}=
\!\frac{p_i\left| \mathbf{g}_{bi}^{T}\mathbf{\Theta \Phi g}_{ai} \right|^2}{\sum\limits_{j=1,j\not =i}^K{\!\left( p_j\left| \mathbf{g}_{bi}^{T}\mathbf{\Theta \Phi g}_{aj} \right|^2 \!\right)}\!+\!|\mathbf{g}_{bi}^{T}\mathbf{\Theta \Phi G}_a\sqrt{\mathbf{\Lambda }}\bm{\eta }_t|^2\!+\!\varUpsilon _{ri}\!+\!\sigma _{i}^{2}}.
\end{align}
\vspace{-0.2cm}
\normalsize
\vspace{-0.4cm}
\end{figure*}

\begin{figure*}[b]
\hrulefill
\vspace{-0.15cm}
\setcounter{equation}{12}
\begin{align}\label{Rif}
R_i\approx\!\log _2\!\left(\!1\!+\frac{p_i\alpha _{bi}\alpha _{ai}\frac{\varepsilon _i\beta _i\tilde{\Gamma}_{i,i}+L\left( \varepsilon _i+\beta _i \right) +L}{\left( \varepsilon _i+1 \right) \left( \beta _i+1 \right)}}{\left( 1+\kappa _r \right) \left( 1+\kappa _t \right) \sum\limits_{j=1}^K{\left( p_j\alpha _{bi}\alpha _{aj}\frac{\varepsilon _i\beta _j\tilde{\Gamma}_{i,j}+L\left( \varepsilon _i+\beta _j \right) +L}{\left( \varepsilon _i+1 \right) \left( \beta _j+1 \right)} \right)}-p_i\alpha _{bi}\alpha _{ai}\frac{\varepsilon _i\beta _i\tilde{\Gamma}_{i,i}+L\left( \varepsilon _i+\beta _i \right) +L}{\left( \varepsilon _i+1 \right) \left( \beta _i+1 \right)}+\sigma _{i}^{2}}\right)
\end{align}
\normalsize
%\hrulefill
\vspace {-0.8cm}
\end{figure*}

\begin{figure*}[t]
\vspace{-0.9cm}
\setcounter{equation}{14}
\vspace {-0.3cm}
\begin{equation}\label{Rithi}
\vspace {-0.2cm}
R_i^{\text{N\!-RIS\!-HWIs}}\approx\!\log _2\!\left(\!1\!+ \frac{p_i\alpha _{bi}\alpha _{ai}\frac{\varepsilon _i\beta _i\Gamma_{i,i}+L\left( \varepsilon _i+\beta _i \right) +L}{\left( \varepsilon _i+1 \right) \left( \beta _i+1 \right)}}{\left( 1+\kappa _r \right) \left( 1+\kappa _t \right) \sum\limits_{j=1}^K{\left( p_j\alpha _{bi}\alpha _{aj}\frac{\varepsilon _i\beta _j\Gamma_{i,j}+L\left( \varepsilon _i+\beta _j \right) +L}{\left( \varepsilon _i+1 \right) \left( \beta _j+1 \right)} \right)}-p_i\alpha _{bi}\alpha _{ai}\frac{\varepsilon _i\beta _i\Gamma_{i,i}+L\left( \varepsilon _i+\beta _i \right) +L}{\left( \varepsilon _i+1 \right) \left( \beta _i+1 \right)}+\sigma _{i}^{2}}\right)
\end{equation}
\vspace{-0.6cm}
\normalsize
\vspace{-0.15cm}
\hrulefill
\end{figure*}

\vspace{-0.4cm}
\section{Introduction}

Reconfigurable intelligent surface (RIS) is a new transmission technology that can configure the radio channel in a desired manner by optimizing the phase shift of each reflecting element \cite{2020Reconfigurable}.
Due to their appealing properties of low power consumption and low cost, RIS-aided wireless communications have attracted much research attentions \cite{200605147,Shen2019Secrecy,2019arXiv190804863P,Tong2020Latency}.
Some initial attempts to study RIS-aided communication systems include RIS-aided full-duplex systems \cite{200605147}, RIS-aided physical layer security \cite{Shen2019Secrecy}, RIS-aided wireless power transfer \cite{2019arXiv190804863P}, RIS-aided mobile edge computing \cite{Tong2020Latency}, and RIS-aided multiuser transmission \cite{9211520}. The cascaded channel estimation was studied in \cite{8937491}.

D2D communication has
received great attention to meet the rapidly increasing
demand for data traffic \cite{8673768}. In D2D communication in
cellular networks, cellular and D2D users transmit signals
simultaneously using the same spectrum of cellular users.
However, there is a paucity of contributions on RIS-aided D2D communications.
In \cite{9366346}, the authors proposed to deploy an RIS that can  assist pairs of devices in their communication when the direct links are blocked by high buildings, plants and walls.

However, most of the existing contributions on D2D communication are based on the assumption of perfect hardware in the radio frequency chains \cite{6525429} and the RISs \cite{9366346}.
%In practice, the hardware of the transceiver can not be ideal because of I/Q imbalance, phase noise, and amplifier nonlinearities \cite{6525429}.
Existing studies have shown that hardware impairments adversely affect the system performance of the multiple-input multiple-output systems \cite{7762840,2021arXiv210205333P}.
In addition, an RIS has low hardware cost \cite{200605147,Shen2019Secrecy,2019arXiv190804863P,Tong2020Latency,9366346}, which is prone to hardware imperfections.

\begin{figure}[t]
\vspace{-0.5cm}
\centering
\includegraphics[scale=1.0]{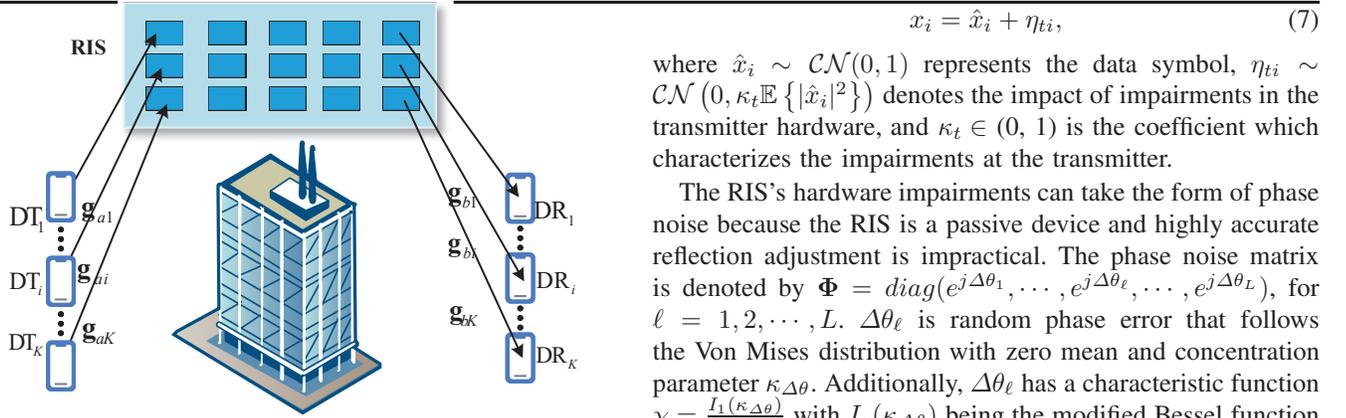}
\vspace{-0.8cm}
\caption{System model for RIS-aided D2D communications.}
\label{multipair}
\vspace{-0.4cm}
\end{figure}

Based on above, a natural question is whether we can use non-ideal low-cost hardware to assist the D2D communications.
Specifically, our contributions are summarized as follows:
1) We consider an RIS-aided D2D communication system over Rician fading channels, considering hardware impairments
both at the terminals and at the RISs.
We assume that the hardware impairment is coupling with the transmission signal at the transceiver.
Moreover, the random phase error caused by the phase noise at the RIS follows the Von Mises distribution;
2) we present the closed-form expressions for the achievable rate in the general case of both hardware impairments both at the terminals and at the RISs, which is then maximized by a genetic algorithm (GA) method. To obtain more design insights, we also study  the two special cases of  no RIS hardware impairments (N-RIS-HWIs) and no transceiver hardware impairments (N-T-HWIs);
3) we provide simulation results to verify our derived expressions. In addition, the exhaustive search method is used to show that our results can achieve the globally optimal solution.

The rest of the letter is organized as follows.
The model for RIS-aided D2D communication system with hardware impairments is depicted in Section
\uppercase\expandafter{\romannumeral2}.
We derive the sum achievable rate's expression and maximize it
in Section \uppercase\expandafter{\romannumeral3}.
Numerical results are presented in Section \uppercase\expandafter{\romannumeral4}. And
conclusions are drawn in Section \uppercase\expandafter{\romannumeral5}.

\section{System Model}

We consider an RIS-aided D2D communication system, where
an RIS assists $K$ pairs of devices to exchange information, as shown in Fig. \ref{multipair}.
%We assume that the base station is aware of all the channel state information (CSI) and performs the optimization of RIS's phase shift with wireless control links.
It is assumed that there are a total of $K$ D2D links in the network, where the $i$th single-antenna transmitter (receiver) is denoted by  $\text {DT}_{i}$ ($\text {DR}_{i}$) for $i = 1,\cdots, K$.
In the overlay mode, the radio resources occupied by D2D links are orthogonal to that of the cellular links, which ensures no interference between D2D links and cellular links.
The RIS includes $L$ scattered-reflection elements. And the phase shift matrix is denoted by $\bm{\Theta} = diag(e^{j\theta_1},\cdots,e^{j\theta_\ell},\cdots,e^{j\theta_L})$, where $\theta_\ell$ is the phase shift of the $\ell$th scattered-reflection element.

The channel between $\text {DT}_{i}$ and the RIS (the RIS and $\text {DR}_{i}$) can be written as
\setcounter{equation}{0}
\vspace{-0.4cm}
\begin{equation}
\vspace{-0.1cm}
\textbf{g}_{ai} =\sqrt{\alpha_{ai}}\textbf{h}_{ai},
\end{equation}
\begin{equation}
\vspace{-0.2cm}
\textbf{g}_{bi} =\sqrt{\alpha_{bi}}\textbf{h}_{bi},
\end{equation}
where $\alpha_{ai}$ and $\alpha_{bi}$ denote the large-scale fading coefficients, and
$\textbf{h}_{ai} \in {{\mathbb C}^{L \times 1}}$ and $\textbf{h}_{bi} \in {{\mathbb C}^{L \times 1}}$ denote
Rician fading channels, which can be expressed as
\vspace{-0.2cm}
\begin{equation}\label{gi}
\vspace{-0.1cm}
\textbf{h}_{ai} =\sqrt{\frac{\varepsilon_i}{\varepsilon_i+1}}\overline{\textbf{h}}_{ai}+\sqrt{\frac{1}{\varepsilon_i+1}} \tilde{\textbf{h}}_{ai},
\end{equation}
\begin{equation}\label{hi}
\vspace{-0.2cm}
\textbf{h}_{bi} =\sqrt{\frac{\beta_i}{\beta_i+1}}\overline{\textbf{h}}_{bi}+\sqrt{\frac{1}{\beta_i+1}} \tilde{\textbf{h}}_{bi},
\end{equation}
where $\varepsilon_i$ and $\beta_i$ denote the Rician factors,
$\tilde{\textbf{h}}_{ai} \in {{\mathbb C}^{L \times 1}}$ and $\tilde{\textbf{h}}_{bi} \in {{\mathbb C}^{L \times 1}}$ are non-line-of-sight components,
whose entries are standard Gaussian random variables with independent and identical distribution,
i.e., $\mathcal{CN}(0,1)$,
and
$\overline{\textbf{h}}_{ai} \in {{\mathbb C}^{L \times 1}}$ and $\overline{\textbf{h}}_{bi} \in {{\mathbb C}^{L \times 1}}$ are line-of-sight (LoS) components. Particularly, $\overline{\textbf{h}}_{ai}$ and $\overline{\textbf{h}}_{bi}$ can be written as
\vspace{-0.1cm}
\begin{equation}
\vspace{-0.3cm}
\!\!\!\overline{\mathbf{h}}_{ai}\left( \varphi _{i}^{a},\varphi _{i}^{e} \right) \!\!=\!\! \left[\!\! \begin{array}{c}\!\!\!\!\!
	1,\cdots ,e^{j2\pi \frac{d}{\lambda}\left( x\sin \varphi _{i}^{a}\sin \varphi _{i}^{e}+y\cos \varphi _{i}^{e} \right)},\cdots ,\\
	e^{j2\pi \frac{d}{\lambda}\left( \left( \sqrt{L}-1 \right) \sin \varphi _{i}^{a}\sin \varphi _{i}^{e}+\left( \sqrt{L}-1 \right) \cos \varphi _{i}^{e} \right)}\\
\end{array} \!\!\!\!\!\right]^T\!\!\!\!\!,
\end{equation}
\begin{equation}
\vspace{-0.2cm}
\overline{\mathbf{h}}_{bi}\left( \varsigma _{i}^{a},\varsigma _{i}^{e} \right) \!\!=\!\! \left[ \begin{array}{c}\!\!\!\!\!
	1,\cdots ,e^{j2\pi \frac{d}{\lambda}\left( x\sin \varsigma _{i}^{a}\sin \varsigma _{i}^{e}+y\cos \varsigma _{i}^{e} \right)},\cdots ,\\
	e^{j2\pi \frac{d}{\lambda}\left( \left( \sqrt{L}-1 \right) \sin \varsigma _{i}^{a}\sin \varsigma _{i}^{e}+\left( \sqrt{L}-1 \right) \cos \varsigma _{i}^{e} \right)}\\
\end{array} \!\!\!\!\!\right]^T\!\!\!\!\!,
\end{equation}
where $0\leqslant x,y\leqslant \sqrt{L}-1$,
$\varphi _{i}^{a}, \varphi _{i}^{e} (\varsigma _{i}^{a}, \varsigma _{i}^{e})$ respectively denote the $i$th pair of devices' azimuth and elevation angles of arrival (angle of departure).
We assume $d=\frac{\lambda}{2}$ in our letter \cite{2020Reconfigurable}.

The signal transmitted from $\text {DT}_{i}$ is given by \cite{7762840}
\vspace{-0.2cm}
\begin{equation}
\vspace{-0.2cm}
x_i=\hat{x}_i+\eta _{ti},
\end{equation}
where $\hat{x}_i$ $\sim \mathcal{CN}(0,1)$ represents the data symbol, $\eta _{ti} \sim \mathcal{C}\mathcal{N}\left( 0,\kappa _t\mathbb{E}\left\{ |\hat{x}_i|^2 \right\} \right) $
denotes the impact of impairments in the transmitter hardware, and $\kappa _t \in$ (0, 1) is the coefficient which characterizes the impairments at the transmitter.

The RIS's hardware impairments can take the form of phase noise because the RIS is a passive device and highly accurate reflection adjustment is impractical.
The phase noise matrix is denoted by $\mathbf{\Phi } = diag(e^{j\varDelta\theta_1},\cdots,e^{j\varDelta\theta_\ell},\cdots,e^{j\varDelta\theta_L})$, for $\ell = 1, 2,\cdots, L$. $\varDelta\theta_\ell$ is random phase error that follows the Von Mises distribution with zero mean and concentration parameter $\kappa_{\varDelta\theta}$. Additionally, $\varDelta\theta_\ell$ has a characteristic function $\chi = \frac{I_1(\kappa _{\varDelta \theta})}{I_0(\kappa _{\varDelta \theta})}$ with $I_p(\kappa _{\varDelta \theta})$ being the modified Bessel function of the first kind and order $p$ \cite{2021arXiv210205333P}.

%\subsection{perfect CSI}
%{\color{blue}We exploit statistical channel state information (CSI) since it can be readily obtained.
%The statistical CSI can be readily obtained since it varies much slowly than the instantaneous CSI.}
We exploit statistical channel state information (CSI), which varies much slowly than the instantaneous CSI and can be readily obtained.
The signal received at $\text {DR}_{i}$ is given by
\vspace{-.25cm}
\begin{align}\label{y_i1}
y_i&=\mathbf{g}_{bi}^{T}\mathbf{\Theta \Phi }\sum_{j=1}^K{\sqrt{p_j}}\mathbf{g}_{aj}x_j+\eta _{ri}+n_i
\nonumber\\
&=\mathbf{g}_{bi}^{T}\mathbf{\Theta \Phi G}_a\sqrt{\mathbf{\Lambda }}\mathbf{x}+\eta _{ri}+n_i
,
\end{align}
%\end{figure*}
where $\mathbf{G}_a\triangleq \left[ \mathbf{g}_{a1},\cdots,\mathbf{g}_{aK} \right]
$, $\mathbf{x}\triangleq \left[ x_1,\cdots,x_K \right] ^{\text{T}}
$, $p_{j}$ denotes the transmit power of $\text U_{Aj}$ with $\mathbf{\Lambda }\triangleq \text{diag}\left( p_1,\cdots,p_K \right)$,
$\eta _{ri}\sim \mathcal{C}\mathcal{N}\left( 0,\varUpsilon _{ri} \right)
$ denotes the impact of impairments in the receiver hardware with $
\varUpsilon _{ri}=\kappa _r\mathbb{E}\left\{ |\mathbf{g}_{bi}^{T}\mathbf{\Theta \Phi G}_a\sqrt{\mathbf{\Lambda }}\mathbf{x}|^2 \right\}
$, $\kappa _r \in$ (0, 1) is the coefficient which characterizes the impairments at the receiver, and $n_{i}$ $\sim \mathcal{CN}(0,\sigma_{i}^2)$ is the additive white Gaussian noise at $\text {DR}_{i}$, for $i = 1,\cdots,K$.

To analyze the achievable rates, we consider the expansion
of \eqref{y_i1} as follows
\vspace{-0.5cm}
\begin{align}\label{y_i2}
y_i&=\sqrt{p_i}\mathbf{g}_{bi}^{T}\mathbf{\Theta \Phi  g}_{ai}\hat{x}_i+\sum_{j=1,j\not =i}^K{\sqrt{p_j}}\mathbf{g}_{bi}^{T}\mathbf{\Theta \Phi g}_{aj}\hat{x}_j
\nonumber\\
&+\mathbf{g}_{bi}^{T}\mathbf{\Theta \Phi G}_a\sqrt{\mathbf{\Lambda }}\bm{\eta}_t+\eta_{ri}+n_i,
\end{align}
where $\bm{\eta }_t\triangleq \left[ \eta _{t1},\cdots,\eta _{tK} \right] ^{T}$.

From \eqref{y_i2}, we obtain the signal-to-interference plus noise ratio at $\text {DR}_{i}$ in \eqref{SINR_i} shown at the bottom of this page.

Therefore, the achievable rate for $\text {DR}_{i}$ can be expressed as
\setcounter{equation} {10}
\vspace{-0.3cm}
\begin{equation}\label{R_i1}
\vspace{-0.2cm}
R_{i}= \mathbb{E}\{\text{log}_2(1+{\rm{\gamma}}_{i})\},
\end{equation}
and the sum achievable rate can be written as
\vspace{-0.1cm}
\begin{equation}\label{C}
\vspace{-0.1cm}
C=\sum_{i=1}^K R_{i}.
\end{equation}

\section{Achievable Rate Analysis}
In this section, we first derive an approximate expression for the sum achievable rate in the following theorem.
Then, we consider two special cases of N-RIS-HWIs and N-T-HWIs.
Finally, we maximize the achievable rate with GA (genetic algorithm) method.

\begin{theorem}\label{theorem1}
In a D2D communication system aided by RIS, the ergodic sum achievable rate of $\text {DR}_{i}$ can be approximated as \eqref{Rif},
where $\tilde{\Gamma}_{i,i}$ and $\tilde{\Gamma}_{i,j}$ can be given by
\setcounter{equation}{13}
\begin{align}\label{toij0}
\tilde{\Gamma}_{i,h}\!=&\mathbb{E}\left\{ \!\left| \sum_{\ell =1}^L{e^{j\left[ \left( \theta _{\ell}+\varDelta \theta _{\ell} \right) +\pi T_{i,h}^{n,m} \right]}} \right|^2\! \right\}\!\!
\nonumber\\
=&L+2\chi ^2\!\!\!\!\!\!\!\!\sum_{1\le m<n\le L}{\!}\!\!\!\!\!\!\!\cos \left( \theta _n\!-\!\theta _m+\pi T^{n,m}_{i,h} \right),
h\in \left\{ i,j \right\}
\end{align}
where $T^{n,m}_{i,h} = \left( x_n-x_m \right) p_{i,h}+\left( y_n-y_m \right) q_{i,h} $,
with $x_z=\lfloor \frac{\left( z-1 \right)}{\sqrt{L}} \rfloor$, $y_z=\left( z-1 \right) mod\sqrt{L}$, $z\in \left\{ m,n \right\}$, $p_{i,h}=\sin \varphi _{i}^{a}\sin \varphi _{i}^{e}+\sin \varsigma _{h}^{a}\sin \varsigma _{h}^{e}$, and     $q_{i,h}=\cos \varphi _{i}^{e}+\cos \varsigma _{h}^{e}$.

\end{theorem}

\itshape \textbf{Proof:}
See Appendix A.
\upshape
\hfill $\Box$

\begin{corollary}\label{cor-1}
Assuming that the RIS hardware is ideal and thus there is no phase noise, i.e.,
$\varDelta\theta_\ell = 0$, for $i = 1, 2, \cdots, N$. It follows $\mathbf{\Phi } = \mathbf{I}_L$.
The rate of $\text {DR}_{i}$ on N-RIS-HWIs can be approximated as \eqref{Rithi},
where $\Gamma_{i,i}$ and $\Gamma_{i,j}$ can be defined as
\vspace{-0.1cm}
\setcounter{equation}{15}
\begin{equation}\label{oij1}
\vspace{-0.1cm}
\Gamma_{i,h}
\!=\!L+2\sum_{\mathclap{1\leq m<n\leq L}}
\cos \left( \theta _n\!-\theta _m+\pi T^{n,m}_{i,h} \right)
, h\in \left\{ i,j \right\}.
\end{equation}

\itshape \textbf{Proof:}
\upshape
When $\mathbf{\Phi } = \mathbf{I}_L$,
we can use \cite[Eq. (20) Eq. (21)]{9366346}, and write
$\mathbb{E}\left\{\left|\textbf{h}_{bi}^T\bm{\Theta }\textbf{h}_{aj}\right|^2\right\}$ and $\mathbb{E}\left\{\left|\textbf{h}_{bi}^T\bm{\Theta }\textbf{h}_{ai}\right|^2\right\}$ as
\vspace{-.2cm}
\begin{equation}\label{E-htheta-ij}
\vspace{-.3cm}
\mathbb{E}\left\{\left|\textbf{h}_{bi}^T\bm{\Theta}\textbf{h}_{ah}\right|^2\right\}
=\frac{\varepsilon_i\beta_h\Gamma_{i,h}+L(\varepsilon_i+\beta_h)+L}{(\varepsilon_i+1)(\beta_h+1)}
, h\in \left\{ i,j \right\},
\end{equation}
where $\Gamma_{i,h}$ is defined in \eqref{oij1}.

By substituting \eqref{E-htheta-ij} into \eqref{Ri3}, we are able to obtain the final result in \eqref{Rithi}.
This completes the proof.
\hfill $\Box$
\end{corollary}

We aim to solve the achievable rate maximization problem by optimizing the phase shifts matrix $\bm{\Theta}$ in the special case with only one pair of users, i.e., $K = 1$. Without loss of generality, the user pair is referred to as user pair $i$.
We can rewrite the achievable rate as
\begin{equation*}
\!\!\!\!\!\!\!\!\!\!\!\!\!\!\!\!\!\!\!\!\!\!\!\!\!\!\!\!\!\!\!\!
\!\!\!\!\!\!\!\!\!\!\!\!\!\!\!\!\!\!\!\!\!\!\!\!\!\!\!\!\!\!\!\!
\!\!\!\!\!\!\!\!\!\!\!\!\!\!\!\!\!\!\!\!\!\!\!\!\!\!\!\!\!\!\!\!
\!\!\!\!\!\!\!\!\!\!\!\!\!\!\!\!\!\!\!\!\!\!\!
R_{i}^{\text{N-RIS-HWIs}}\approx
\end{equation*}
\vspace{-0.8cm}
\begin{equation}
\vspace{-0.2cm}
\!\log _2\!\!\left( \!\!\!1\!+\!\frac{\frac{p_i\alpha _{bi}\alpha _{ai}}{\left( \varepsilon _i+1 \right) \left( \beta _i+1 \right)}\left( \varepsilon _i\beta _i\Gamma _{i,i}\!+\!L\left(\! \varepsilon _i+\beta _i \right) \!+\!L \right)}{\!\!\!\!\left( \kappa _t\kappa _r\!+\!\kappa _t\!+\!\kappa _r \!\right)\!\! \frac{p_i\alpha _{bi}\alpha _{ai}}{\left( \varepsilon _i+1 \right) \left( \beta _i+1 \right)}\!\!\left(\! \varepsilon _i\beta _i\Gamma _{i,i}\!+\!L\!\left( \varepsilon _i\!+\!\beta _i \right) \!+\!L\! \right) \!+\!\sigma _{i}^{2}} \!\!\right)\!\!.
\end{equation}
The rate depends on the phase shifts matrix $\bm{\Theta}$ only through the intermediate variable $\Gamma _{i,i}$. Since $\Gamma _{i,i}\!=\left| \sum_{\ell =1}^L{e^{j\left[ \theta _{\ell}+\pi T_{i,h}^{n,m} \right]}} \right|^2\leqslant \left( \sum_{\ell =1}^L{|e^{j\left[ \theta _{\ell}+\pi T_{i,i}^{n,m} \right]}|} \right) ^2=L^2$
and $\Gamma _{i,i} \gg 0$,
the optimization problem
can be formulated as follows
\vspace{-0.5cm}
\begin{subequations}
\begin{alignat}{2}
\max_{\Gamma _{i,i}} \quad & R_{i}^{\text{N-RIS-HWIs}} \\
\mbox{s.t.}\quad
&0\leqslant \Gamma _{i,i}\leqslant L^2.
\end{alignat}
\end{subequations}
The expression for the first-order derivative of $\text{SINR}_i\left( \Gamma _{i,i} \right)$ with respect to $\Gamma _{i,i}$ is
\vspace{-.05cm}
\begin{equation*}
\vspace{-.8cm}
\!\!\!\!\!\!\!\!\!\!\!\!\!\!\!\!\!\!\!\!\!\!\!\!\!\!\!\!\!\!\!\!\!\!\!\!\!\!\!\!\!\!\!\!\!\!\!\!\!\!
\!\!\!\!\!\!\!\!\!\!\!\!\!\!\!\!\!\!\!\!\!\!\!\!\!\!\!\!\!\!\!\!\!\!\!\!\!\!\!\!\!\!\!\!\!\!\!\!\!\!
\frac{\partial \text{SINR}_i\left( \Gamma_{i,i} \right)}{\partial \Gamma_{i,i}}=
\end{equation*}
\begin{align}
&\frac{\frac{p_i\alpha _{bi}\alpha _{ai}}{\left( \varepsilon _i+1 \right) \left( \beta _i+1 \right)}\varepsilon _i\beta _i\sigma _{i}^{2}}{\left[ \!\left(\! \kappa _t\kappa _r\!+\!\kappa _t\!+\!\kappa _r \!\right)\! \frac{p_i\alpha _{bi}\alpha _{ai}}{\left( \varepsilon _i+1 \right) \left( \beta _i+1 \right)}\left( \varepsilon _i\beta _i\Gamma _{i,i}\!+\!L\left( \varepsilon _i+\beta _i \right) \!+\!L \right) \!+\!\sigma _{i}^{2} \right] ^2}
\nonumber\\
&\geqslant 0.
\end{align}
Thus, the optimal design for $\bm{\Theta}$ corresponds to setting $\Gamma_{i,i} = L^2$, where the optimal phase shifts of all the RIS elements are given by $\theta _{\ell}=-\pi T_{i,h}^{n,m}+C_0, \forall \ell$, and $C_0$ is an arbitrary constant.

Considering this single-user system with optimal phase shift, we assume the transmit power is scaled as $p_i = \frac{E_u}{L^2}$ and $p_i = \frac{E_u}{L}$ with $L\rightarrow \infty$.
If the transmit power is scaled as $p_i = \frac{E_u}{L^2}$, the rate becomes
\begin{align}
&R_{i}^{\text{N}-\text{RIS}-\text{HWIs}}\rightarrow
\\ \nonumber
&\log _2\!\!\left( 1\!+\!\frac{\frac{E_u\alpha _{bi}\alpha _{ai}\varepsilon _i\beta _i}{\left( \varepsilon _i+1 \right) \left( \beta _i+1 \right)}}{\!\!\!\!\left( \kappa _t\kappa _r\!+\!\kappa _t\!+\!\kappa _r\! \right) \frac{E_u\alpha _{bi}\alpha _{ai}\varepsilon _i\beta _i}{\left( \varepsilon _i+1 \right) \left( \beta _i+1 \right)}\,\,\!\!\!+\!\sigma _{i}^{2}}\!\! \right) ,\ \text{as}\  L\rightarrow \infty.
\end{align}
If the transmit power is scaled as $p_i = \frac{E_u}{L}$, the rate becomes
\vspace{-.2cm}
\begin{equation}
\vspace{-.2cm}
R_{i}^{\text{N}-\text{RIS}-\text{HWIs}}\rightarrow \!\log _2\!\!\left( 1\!+\!\frac{1}{\kappa _t\kappa _r\!+\!\kappa _t\!+\!\kappa _r\,\,\!\!\,\, \!\!\!} \right),\; \text{as}\;  L\rightarrow \infty.
\end{equation}

In this case, the achievable rate only depends on the impairment coefficients at the transceiver
when $L$ grows into infinity.

\begin{corollary}\label{cor-2}
Assuming that the transceiver hardware is ideal, i.e., $\kappa _t = \kappa _r = 0$.
The rate of $\text {DR}_{i}$ with N-T-HWIs can be approximated as
\begin{align}\label{Rirhi}
&R_i^{\text{N-T-HWIs}}\approx
\nonumber\\
&\log _2\!\left(\!1\!+\! \frac{p_i\alpha _{bi}\alpha _{ai}\frac{\varepsilon _i\beta _i\tilde{\Gamma}_{i,i}+L\left( \varepsilon _i+\beta _i \right) +L}{\left( \varepsilon _i+1 \right) \left( \beta _i+1 \right)}}{ \!\!\!\!\!\!\!\sum\limits_{j=1,j \not =i}^K{\left( p_j\alpha _{bi}\alpha _{aj}\frac{\varepsilon _i\beta _j\tilde{\Gamma}_{i,j}+L\left( \varepsilon _i+\beta _j \right) +L}{\left( \varepsilon _i+1 \right) \left( \beta _j+1 \right)} \right)}\!+\!\sigma _{i}^{2}} \!\right)\!.
\end{align}
\end{corollary}

We design the phase shift with one pair of users, the achievable rate is
\vspace{-.3cm}
\begin{equation}
\vspace{-.25cm}
R_{i}^{\text{N-T-HWIs}}\approx \log _2\!\left( \!1\!+\!\frac{p_i\alpha _{bi}\alpha _{ai}\frac{\varepsilon _i\beta _i\tilde{\Gamma}_{i,i}+L\left( \varepsilon _i+\beta _i \right) +L}{\left( \varepsilon _i+1 \right) \left( \beta _i+1 \right)}}{\!\!\sigma _{i}^{2}}\! \right),
\end{equation}
which is a monotonically increasing function of $\tilde{\Gamma}_{i,i}$.
The optimal design for $\bm{\Theta}$ corresponds to setting $\tilde{\Gamma}_{i,i} = L^2$, where the optimal phase shifts of all the RIS elements are given by $\theta _{\ell}=-\pi T_{i,h}^{n,m}+C_1, \forall \ell$, and $C_1$ is an arbitrary constant.

Considering this single-user system with optimal phase shift, we assume the transmit power is scaled as $p_i = \frac{E_u}{L^2}$ with $L\rightarrow \infty$.
If the transmit power is scaled as $p_i = \frac{E_u}{L^2}$, the rate becomes
\vspace{-0.3cm}
\begin{equation}
\vspace{-0.2cm}
R_{i}^{\text{N}-\text{T}-\text{HWIs}}\!\!\!\!\rightarrow\! \log _2\!\!\left( \!\!1\!+\!\frac{E_u\alpha _{bi}\alpha _{ai}\varepsilon _i\beta _i}{\!\!\sigma _{i}^{2}\left( \varepsilon _i+1 \right) \left( \beta _i+1 \right)}\! \right)\!\!,\; \text{as}\;  L\rightarrow \infty.
\end{equation}

Due to the first-order derivative of the sum rate with respect to the phase shift $\bm{\Theta}$ is quite hard to obtain,
we maximize the rate with the GA method adopted in \cite{9366346} by optimizing the phase shifts.
The complexity of the proposed GA algorithm is $N_t*n$, where $N_t$ is the population size, and $n$ is the number of generations evaluated.
We take into account both continuous phase shifts (CPSs) and discrete phase shifts (DPSs).
In practice, only a limited number of phase shifts can be used.
We assume that the phase shift of the reflecting elements is quantized  with $B$ bits  when considering DPS, and thus $2^B$ phase shifts can be chosen for each reflecting element.

\section{ Numerical Results}

We evaluate the impact of various parameters on the data rate performance.
We set SNR = $p_i$, $\varepsilon_i$ = 10 dB, $\kappa =\kappa _t = \kappa _r$, $\kappa_{\varDelta\theta}$ = 4, $\sigma_i^2 = 1$, and $\varphi _{i}^{a} = \varphi _{i}^{e}$ and $\varsigma _{i}^{a} = \varsigma _{i}^{e}$
are respectively set as
$\varphi _{i}$ and $\varsigma _{i}$ in \cite{9366346}, for $i = 1, \cdots, K$.
Other parameters are set the same as \cite{9366346}.

\begin{figure}[t]
\vspace{-1.2cm}
\centering
\includegraphics[scale=0.50]{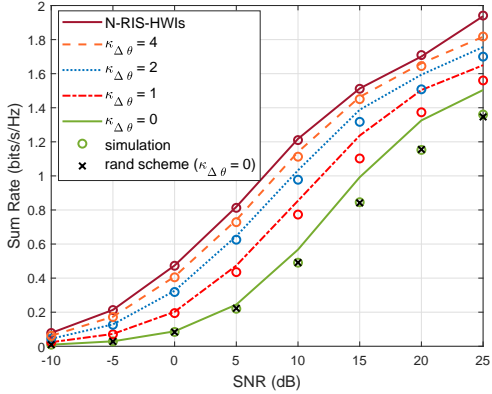}
\vspace{-0.35cm}
      \caption{Sum achievable rate versus SNR with $L$ = 16, $K$ = 6, $\kappa = 0.05$, $B$ = 2.} \label{fig1}
\vspace{-0.4cm}
\end{figure}

In Fig. \ref{fig1}, we depict the rate in \eqref{C} versus the SNR obtained from the approximate expression in \eqref{Ri3} and Monte-Carlo simulations from  \eqref{R_i1}, when $L$ = 16, $K$ = 6, $\kappa$ = 0.05, $B$ = 2.
The Monte-Carlo simulation results match the analytical expressions well, which verifies the data rate performance.
Furthermore, the rate with  N-RIS-HWIs performs better than the case with RIS-HWIs.
As $\kappa_{\varDelta\theta}$ decreases, the rate decreases.
Additionally, when $\kappa_{\varDelta\theta} = 0$, the Von Mises distribution degenerates into the uniform distribution, i.e., $\varDelta\theta_\ell \sim \mathcal{U}[-\pi, \pi)$, for $\ell = 1, 2,\cdots, L$.
In this case, the random scheme curve has the same performance as that of our derived results, which demonstrate that there is no need to optimize the phase shift. The phenomenon is consistent with the result in \cite{2021arXiv210205333P}.

\begin{figure}[t]
\vspace{-0.3cm}
\centering
\includegraphics[scale=0.50]{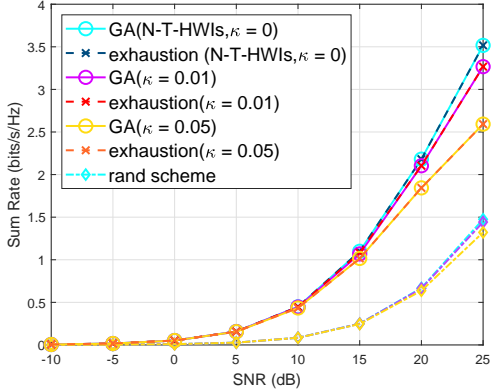}
\vspace{-0.35cm}
      \caption{Sum achievable rate versus SNR with $L$ = 9, $K$ = 2 and $B$ = 2.} \label{fig2}
\vspace{-0.5cm}
\end{figure}
Fig. \ref{fig2} shows the sum rate versus the SNR when $L$ = 8, $K$ = 2, and $B$ = 2.
We can reduce the  harmful impact of the T-HWIs by tuning the phase shifts of the RIS.
Compared with the random scheme, the optimal phase shift can achieve higher rate.
It is worth noting that the proposed GA method achieves similar performance to the exhaustive search, which implies that our proposed algorithm can achieve almost the globally optimal solution.
Moveover, we observe that different levels of hardware impairment obtain different rate:
the higher the level of the hardware impairment, the worse the performance of the rate.
The rate with the ideal transceiver hardware (N-T-HWIs, $\kappa$ = 0) performs the best, as expected.

\begin{figure}[t]
\vspace{-1.2cm}
\centering
\includegraphics[scale=0.50]{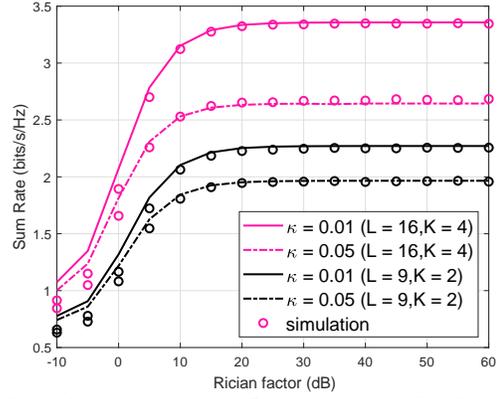}
\vspace{-0.35cm}
      \caption{Sum achievable rate versus Rician factor with $B$ = 2 and SNR = 20 dB.} \label{fig3}
\vspace{-0.4cm}
\end{figure}
Fig. \ref{fig3} illustrates the rate versus the Rician factor when SNR = 20 dB and $B$ = 2.
In all cases, with different values of $L$ and $K$, the rates approach the fixed value as Rician factor $\varepsilon_i \rightarrow \infty$.
This is because the channels are mainly influenced by LoS component when Rician factor is large.

\begin{figure}[t]
\vspace{-0.3cm}
\centering
\includegraphics[scale=0.50]{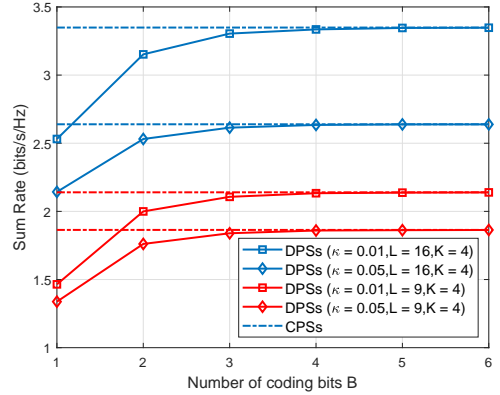}
\vspace{-0.35cm}
      \caption{Sum achievable rate versus $B$ with $\varepsilon_i$ = 10 dB and SNR = 20 dB.} \label{fig4}
\vspace{-0.4cm}
\end{figure}
Fig. \ref{fig4} shows the sum rate versus $B$.
The figure shows that, under the condition of DPS, three quantization bits are sufficient, which provides useful insight into the engineering design of systems aided by RISs. In addition, to resolve the issue caused by imperfect hardware, we can increase the number of low-cost reflection elements.

\section{Conclusion}
In this letter, we investigated an RIS-aided D2D communication system over Rician fading channels, considering hardware impairments both at the terminals and the RISs. We considered two special cases of  N-RIS-HWIs and N-T-HWIs. We derived closed-form expressions for the achievable rate of different cases.
The impact of the hardware impairment on the system performance have been observed.
To resolve
the issue caused by imperfect hardware, we can increase the
number of low-cost reflection elements.
Additionally, three quantization bits are sufficient for the DPSs setup, which provides useful insight into the engineering design of systems aided by RISs.
Moreover, the extension to jointly optimizing power allocation and the phase shifts of the RIS will be left for our future work.
%We derived an approximate \textcolor{blue}{expression for} the achievable rate.
%Based on the derived analytical framework, we developed a GA-based algorithm for optimizing the achievable sum achievable rate, which can be applied to CPS-based and DPS-based implementations.
%%Simulation results verified the correctness of the proposed GA-based method.
%\textcolor{blue}{
%Simulation results verified our derivation and benefit optimization method.}

\begin{appendices}
\section{Proof of Theorem \ref{theorem1} }

\begin{figure*}[ht]
\vspace{-0.6cm}
%\hrulefill
\setcounter{equation}{22}
\vspace{-0.2cm}
\begin{equation}\label{Ri2}
\vspace{-0.4cm}
R_i  \approx\log _2\left( 1+\frac{p_i\mathbb{E}\left\{ \left| \mathbf{g}_{bi}^{T}\mathbf{\Theta \Phi g}_{ai} \right|^2 \right\}}{\sum\limits_{j=1,j\not =i}^K{\left( p_j\mathbb{E}\left\{ \left| \mathbf{g}_{bi}^{T}\mathbf{\Theta \Phi g}_{aj} \right|^2 \right\} \right)}+\mathbb{E}\left\{ |\mathbf{g}_{bi}^{T}\mathbf{\Theta \Phi  G}_a\sqrt{\mathbf{\Lambda }}\bm{\eta }_t|^2 \right\} +\varUpsilon _{ri}+\sigma _{i}^{2}} \right)
\end{equation}
\normalsize
\vspace{-0.15cm}
\end{figure*}

\begin{figure*}[ht]
\hrulefill
\vspace{-0.15cm}
\setcounter{equation}{25}
\begin{align}\label{gamma_1}
&\varUpsilon _{rj}=\kappa _r\mathbb{E}\left\{ |\mathbf{g}_{bi}^{T}\mathbf{\Theta \Phi G}_a\sqrt{\mathbf{\Lambda }}\mathbf{x}|^2 \right\}
%\nonumber\\
%&=\beta _r\mathbb{E}\left\{ \mathbf{g}_{bi}^{T}\mathbf{\Theta \Phi  G}_a\sqrt{\varLambda}\mathbf{xx}^H\sqrt{\varLambda}^H\mathbf{G}^H_a\mathbf{\Phi}^H \mathbf{\Theta}^H   \mathbf{g}_{bi}^{*} \right\}
=\kappa _r\mathbb{E}\left\{ \mathbf{g}_{bi}^{T}\mathbf{\Theta \Phi  G}_a\sqrt{\mathbf{\Lambda }}\left( \mathbf{\hat{x}}+\bm{\eta }_t \right) \left( \mathbf{\hat{x}}^H+\bm{\eta }_t^H \right) \sqrt{\mathbf{\Lambda }}^H\mathbf{G}^H_a\mathbf{\Phi}^H \mathbf{\Theta}^H   \mathbf{g}_{bi}^{*} \right\}
\nonumber\\
&\!\!\!\!\!\!
\overset{\text{(a)}}{=}\!\kappa _r\mathbb{E}\!\left\{\! \mathbf{g}_{bi}^{T}\mathbf{\Theta \Phi G}_a\sqrt{\mathbf{\Lambda }}\mathbb{E}\!\left\{\! \mathbf{\hat{x}\hat{x}}^H\!+\!\mathbf{\hat{x}}\bm{\eta }_t^H\!+\!\bm{\eta }_t\mathbf{\hat{x}}^H\!+\!\bm{\eta }_t\bm{\eta }_t^H \!\right\}\! \sqrt{\mathbf{\Lambda }}^H\mathbf{G}^H_a\mathbf{\Phi}^H \mathbf{\Theta}^H   \mathbf{g}_{bi}^{*} \!\right\}\!
\overset{\text{(b)}}{=}\!\kappa _r\left( 1\!+\!\kappa _t \right) \mathbb{E}\!\left\{ \mathbf{g}_{bi}^{T}\mathbf{\Theta \Phi G}_a\mathbf{\Lambda G}^H_a\mathbf{\Phi}^H \mathbf{\Theta}^H   \mathbf{g}_{bi}^{*} \!\right\}\!
\end{align}
\vspace{-0.4cm}
\normalsize
\vspace{-0.45cm}
\end{figure*}

\begin{figure*}[ht]
\hrulefill
\vspace{-0.15cm}
\begin{equation*}
\vspace{-0.2cm}
R_i\approx\!\log _2\!\left(\!1\!+ \frac{p_i\alpha _{bi}\alpha _{ai}\mathbb{E}\left\{ \left| \mathbf{h}_{bi}^{T}\mathbf{\Theta \Phi h}_{ai} \right|^2 \right\}}{\sum\limits_{j=1,j\not =i}^K{\left( p_j\alpha _{bi}\alpha _{aj}\mathbb{E}\left\{ \left| \mathbf{h}_{bi}^{T}\mathbf{\Theta \Phi h}_{aj} \right|^2 \right\} \right)}+\left( \kappa _t\kappa _r+\kappa _t+\kappa _r \right) \alpha _{bi}\sum\limits_{j=1}^K{\left( \alpha _{aj}p_j\mathbb{E}\left\{ |\mathbf{h}_{bi}^{T}\mathbf{\Theta \Phi  h}_{aj}|^2 \right\} \right)}+\sigma _{i}^{2}}\right)
\end{equation*}
\setcounter{equation}{29}
\begin{equation}\label{Ri3}
\vspace{-0.25cm}
=\log _2\!\left(\!1\!+\frac{p_i\alpha _{bi}\alpha _{ai}\mathbb{E}\left\{ \left| \mathbf{h}_{bi}^{T}\mathbf{\Theta \Phi h}_{ai} \right|^2 \right\}}{\left( 1+\kappa _r \right) \left( 1+\kappa _t \right) \sum\limits_{j=1}^K{\left( p_j\alpha _{bi}\alpha _{aj}\mathbb{E}\left\{ |\mathbf{h}_{bi}^{T}\mathbf{\Theta \Phi h}_{aj}|^2 \right\} \right)}-p_i\alpha _{bi}\alpha _{ai}\mathbb{E}\left\{ \left| \mathbf{h}_{bi}^{T}\mathbf{\Theta \Phi h}_{ai} \right|^2 \right\} +\sigma _{i}^{2}}\right)
\end{equation}
\vspace{-0.45cm}
\hrulefill
\normalsize
\vspace{-0.25cm}
\end{figure*}

Using Lemma~1 in \cite{zhang2014power}, $R_i$ in \eqref{R_i1} is approximated as \eqref{Ri2}.

Next, we derive $\mathbb{E}\left\{ \left| \mathbf{g}_{bi}^{T}\mathbf{\Theta \Phi g}_{ai} \right|^2 \right\}$, $\mathbb{E}\left\{ \left| \mathbf{g}_{bi}^{T}\mathbf{\Theta \Phi g}_{aj} \right|^2 \right\}$, $\mathbb{E}\left\{ |\mathbf{g}_{bi}^{T}\mathbf{\Theta \Phi G}_a\sqrt{\mathbf{\Lambda }}\bm{\eta }_t|^2 \right\}$, and $\varUpsilon _{rj}$. To begin with, we have
\setcounter{equation}{23}
\vspace{-.2cm}
\begin{equation}\label{hhij}
\vspace{-.2cm}
\mathbb{E}\left\{ \left| \mathbf{g}_{bi}^{T}\mathbf{\Theta \Phi g}_{ah} \right|^2 \right\} =\alpha _{bi}\alpha _{ah}\mathbb{E}\left\{ \left| \mathbf{h}_{bi}^{T}\mathbf{\Theta \Phi h}_{ah} \right|^2 \right\}, h\in \left\{ i,j \right\}.
\end{equation}
Both $\mathbb{E}\left\{ |\mathbf{g}_{bi}^{T}\mathbf{\Theta \Phi G}_a\sqrt{\mathbf{\Lambda }}\bm{\eta }_t|^2 \right\}$ and $\varUpsilon _{rj}$ contain the terms of $\mathbb{E}\left\{ \left| \mathbf{h}_{bi}^{T}\mathbf{\Theta \Phi h}_{ai} \right|^2 \right\}$ and $\mathbb{E}\left\{ \left| \mathbf{h}_{bi}^{T}\mathbf{\Theta \Phi h}_{aj} \right|^2 \right\}$. We derive these items later.

Then, $\mathbb{E}\left\{ |\mathbf{g}_{bi}^{T}\mathbf{\Theta \Phi G}_a\sqrt{\mathbf{\Lambda }}\bm{\eta }_t|^2 \right\}$ can be written as follows
\vspace{-.2cm}
\begin{equation}\label{hard}
\vspace{-.2cm}
\!\!\!\!\!\!\mathbb{E}\!\left\{\! |\mathbf{g}_{bi}^{T}\mathbf{\Theta \Phi G}_a\sqrt{\mathbf{\Lambda }}\bm{\eta }_t|^2 \right\} \!\!=\!\! \kappa _t\mathbb{E}\!\left\{ \mathbf{g}_{bi}^{T}\mathbf{\Theta \Phi G}_a\mathbf{\Lambda} \mathbf{G}^H_a\mathbf{\Phi}^H \mathbf{\Theta}^H   \mathbf{g}_{bi}^{*} \right\}.\!\!\!
\end{equation}
$\varUpsilon _{rj}$ also contains the terms of $\mathbb{E}\left\{ \mathbf{g}_{bi}^{T}\mathbf{\Theta \Phi G}_a\mathbf{\Lambda} \mathbf{G}^H_a\mathbf{\Phi}^H \mathbf{\Theta}^H   \mathbf{g}_{bi}^{*} \right\}$, which we will derive later.

Moreover, $\varUpsilon _{rj}$ is calculated in \eqref{gamma_1} at the top of this page.
Equality in (a) is obtained because $\mathbf{\hat{x}}$ is independent of both $\mathbf{g}_{bi}$ and $\mathbf{G}_a$; $\bm{\eta }_t$ is also independent of both $\mathbf{g}_{bi}$ and $\mathbf{G}_a$.
Additionally, equality in (b) uses the following results
\setcounter{equation} {26}
\vspace{-0.2cm}
\begin{equation}
\vspace{-0.2cm}
\mathbb{E}\left\{ \mathbf{\hat{x}\hat{x}}^H \right\} =\mathbf{I}_K,
\
\mathbb{E}\left\{ \bm{\eta }_t\bm{\eta }_t^H \right\} =\kappa _t\mathbf{I}_K,
\end{equation}
\begin{equation}
\vspace{-0.15cm}
\mathbb{E}\left\{ \mathbf{\hat{x}}\bm{\eta }_t^H \right\} =\mathbf{0},
\
\mathbb{E}\left\{ \bm{\eta }_t\mathbf{\hat{x}}^H \right\} =\mathbf{0}.
\end{equation}

We can find $\mathbb{E}\left\{ |\mathbf{g}_{bi}^{T}\mathbf{\Theta \Phi G}_a\sqrt{\mathbf{\Lambda }}\bm{\eta }_t|^2 \right\}$ and $\varUpsilon _{rj}$ contain the item $\mathbb{E}\left\{ \mathbf{g}_{bi}^{T}\mathbf{\Theta \Phi G}_a\mathbf{\Lambda G}^H_a\mathbf{\Phi}^H \mathbf{\Theta}^H   \mathbf{g}_{bi}^{*}\right\}$, given by
\begin{align*}
&\mathbb{E}\left\{ \mathbf{g}_{bi}^{T}\mathbf{\Theta \Phi G}_a\mathbf{\Lambda G}^H_a\mathbf{\Phi}^H \mathbf{\Theta}^H   \mathbf{g}_{bi}^{*}
\right\}
\nonumber\\
&\overset{\text{(c)}}{=}\alpha _{bi}\mathbb{E}\left\{ \mathbf{h}_{bi}^{T}\mathbf{\Theta \Phi H}_a\sqrt{\mathbf{D}_a}\mathbf{\Lambda }\sqrt{\mathbf{D}_a}^{H}\mathbf{H}_{a}^{H}\mathbf{\Theta \Phi }^{H}\mathbf{h}_{bi}^{*} \right\}
\nonumber\\
&=\alpha _{bi}\mathbb{E}\left\{ \sum_{j=1}^K{\alpha _{aj}p_j|\mathbf{h}_{bi}^{T}\mathbf{\Theta \Phi h}_{aj}|^2} \right\}
\end{align*}
\vspace{-.5cm}
\begin{equation}\label{inner}
\vspace{-.2cm}
\!\!\!\!\!\!\!\!\!\!\!\!\!\!\!\!\!\!\!\!\!\!\!\!\!\!\!\!\!\!\!\!\!\!\!\!\!
=\sum_{j=1}^K{\alpha _{bi} \alpha _{aj}p_j\mathbb{E}\left\{ |\mathbf{h}_{bi}^{T}\mathbf{\Theta \Phi h}_{aj}|^2 \right\} }.
\end{equation}
We obtain equality in (c) by defining $\mathbf{D}_a\triangleq \text{diag}\left( \alpha _{a1},\cdots,\alpha _{aK} \right)$.

Substituting \eqref{hhij}, \eqref{hard}, \eqref{gamma_1} and \eqref{inner} into \eqref{Ri2}, we obtain \eqref{Ri3}.
Employing the results of \cite[Eq. (20) Eq. (21)]{9366346} one obtains
\vspace{-0.1cm}
\setcounter{equation}{30}
\begin{equation}\label{tE-htheta-ii}
\vspace{-0.25cm}
\mathbb{E}\left\{\left|\textbf{h}_{bi}^T\bm{\Theta \Phi}\textbf{h}_{ah}\right|^2\right\}=
\frac{\varepsilon_i\beta_h\tilde{\Gamma}_{i,h}+L(\varepsilon_i+\beta_h)+L}{(\varepsilon_i+1)(\beta_h+1)},
h\in \left\{ i,j \right\},
\end{equation}
where $\tilde{\Gamma}_{i,h}$ is defined by
\vspace{-0.2cm}
\begin{equation}\label{toii0}
\vspace{-0.2cm}
\tilde{\Gamma}_{i,h}\!\!=\!\!L\!+\!2\!\!\!\!\!\!\!\!\sum_{1\le m<n\le L}\!\!\!\!\!\!\!{\mathbb{E}\left\{ \cos \left[ \left( \theta _n+\varDelta \theta _n \right) \!-\left( \!\theta _m+\varDelta \theta _m \right) +\pi T^{n,m}_{i,h} \right] \right\}}.
\end{equation}
With the help of
\vspace{-0.2cm}
\begin{equation*}
\mathbb{E}\left\{ \cos \left[ \left( \theta _n+\varDelta \theta _n \right) \!-\left( \!\theta _m+\varDelta \theta _m \right) +\pi T^{n,m}_{i,h} \right] \right\}
\end{equation*}
%\nonumber\\
%\begin{align*}
%&=\mathbb{E}\left\{ \text{Re}\left( \boldsymbol{e}^{j\left[ \varDelta \theta _n-\varDelta \theta _m \right]} \right) \right\} \cos \left[ \left( \theta _n\!-\theta _m+\pi T^{n,m}_{i,h} \right) \right]
%\nonumber\\
%&-\mathbb{E}\left\{ \text{Im}\left( \boldsymbol{e}^{j\left[ \varDelta \theta _n-\varDelta \theta _m \right]} \right) \right\} \sin \left[ \left( \theta _n\!-\theta _m+\pi T^{n,m}_{i,h} \right) \right]
%\end{align*}
\vspace{-.4cm}
\begin{equation}\label{Ecos}
\vspace{-.2cm}
%\!\!\!\!\!\!\!\!\!\!\!\!\!\!\!\!\!\!\!\!\!\!\!\!\!\!\!\!\!\!\!\!\!\!\!\!\!\!\!\!\!\!\!\!
=\chi ^2\cos \left[ \left( \theta _n\!-\theta _m+\pi T^{n,m}_{i,h} \right) \right],
\end{equation}
$\tilde{\Gamma}_{i,h}$ can be further given in \eqref{toij0}.

By substituting \eqref{tE-htheta-ii} into \eqref{Ri3}, we are able to obtain the final result in \eqref{Rif}.
The proof of \textbf{theorem 1} is completed.
\end{appendices}

%
%%
%%
%%
%%%\section*{Acknowledgment}
%%
%%
%%%The authors would like to thank...
%%
%%
%%
%%\ifCLASSOPTIONcaptionsoff
%%  \newpage
%%\fi

%%\begin{thebibliography}{1}
%%% You can use other form of bib file by changing here...
%%
%%\bibitem{Cui2014Coding}
%%E.~Basar, M.~Di~Renzo, J.~de~Rosny, M.~Debbah, M.-S. Alouini, and R.~Zhang,
%%  ``Coding metamaterials, digital metamaterials and programmable metamaterials,''
%%  \emph{Light-Science \& Applications}, 2019.
%%
%%\end{thebibliography}

%\vspace{-0.2cm}

\bibliographystyle{IEEEtran}
\bibliography{IEEEabrv,Ref}

\iffalse
\begin{IEEEbiography}{Yuguang ``Michael'' Fang}
Biography text here.
\end{IEEEbiography}
\fi

%It is not necessary to upload the biography when you submit your manuscript.

\end{document}